\documentclass[12pt,letterpaper]{article}
\usepackage[a4paper, total={7in, 10in}]{geometry}

\usepackage{graphicx}
\usepackage{helvet}

\usepackage{indentfirst}
\usepackage{authblk}

\usepackage{hyperref}
\usepackage{amsmath} 
\usepackage{amssymb} 
\usepackage{orcidlink}
\usepackage{setspace}
\usepackage{natbib}
\setcitestyle{numbers,square,comma,super}

\usepackage{pdfpages}
\newcommand{\journalname}{The Innovation}
\newcommand{\journaldetails}{2026, Volume 7, Issue 3}

\makeatletter

\newcommand{\Rmnum}[1]{\expandafter\@slowromancap\romannumeral #1@}
\renewcommand\@alph[1]{\number #1}
\makeatother

\makeatletter
\renewcommand{\maketitle}{\bgroup\setlength{\parindent}{0pt}
\begin{flushleft}
  {\LARGE \textbf{\@title}}
  
  \vspace{1cm}
  
  \@author
  
  \vspace{0.5cm}
  
  \scriptsize
  Received: July 14, 2025; Accepted: October 27, 2025; \href{https://doi.org/10.1016/j.xinn.2025.101169}{https://doi.org/10.1016/j.xinn.2025.101169}
  
  \textit{\journalname} \journaldetails
  
\end{flushleft}\egroup}
\makeatother

\title{The Faintest, Extremely Variable X-ray Tidal Disruption Event from a Supermassive Black Hole Binary?}

\author[1,2]{Mengqiu Huang,}
\author[1,2]{ Yongquan Xue*,}
\author[3]{ Shuo Li**,}

\author[4,5]{ Fukun Liu,}
\author[1,2]{ Shifu Zhu,}

\author[6]{ Jin-Hong Chen,}
\author[7,8]{ Rong-Feng Shen,}

\author[1,2]{ Yibo Wang,}
\author[9,10]{ Yi Yang,}

\author[1,2]{ Ning Jiang,}

\author[11,12,13]{ Franz Erik Bauer,}
\author[14,15]{ Cristian Vignali,}
\author[16]{ Fan Zou,}
\author[1,2]{ Jialai Wang,}

\author[10]{ Alexei V. Filippenko,}
\author[17,18]{ Bin Luo,}
\author[1,2]{ Chen Qin,}
\author[19]{ Jonathan Quirola-V\'asquez,}
\author[1,2]{ Jun-Xian Wang,}
\author[1,2]{ Lulu Fan,}
\author[20]{ Mouyuan Sun,}
\author[21]{ Qingwen Wu,}
\author[22]{ Qingling Ni,}
\author[10]{ Thomas G. Brink,}
\author[1,2]{ Tinggui Wang,}
\author[10]{ Weikang Zheng,}
\author[23]{ Xinwen Shu,}
\author[24]{ Xuechen Zheng,}
\author[1,2]{ Xiaozhi Lin,}
\author[1,2]{ Xu Kong,}
\author[17,18]{ Yijun Wang,}
\author[1,2]{ Yibin Luo,}
\author[1,2]{ Zheyu Lin}

\affil[1]{Department of Astronomy, University of Science and Technology of China, Hefei, 230026, Anhui, China}

\affil[2]{School of Astronomy and Space Science, University of Science and Technology of China, Hefei, 230026, Anhui, China}

\affil[3]{National Astronomical Observatories, Chinese Academy of Sciences, Beijing, 100012, China}

\affil[4]{Department of Astronomy, School of Physics, Peking University, Beijing, 100871, China}

\affil[5]{Kavli Institute for Astronomy and Astrophysics, Peking University, Beijing, 100871, China}
            
\affil[6]{Department of Physics, University of Hong Kong, Pokfulam Road, Hong Kong, China}

\affil[7]{School of Physics and Astronomy, Sun Yat-sen University, Zhuhai, 519082, China}

\affil[8]{CSST Science Center for the Guangdong-Hong Kong-Macau Greater Bay Area, Sun Yat-sen University, Zhuhai, 519082, China}

\affil[9]{Department of Physics, Tsinghua University, Beijing, 100084, China}

\affil[10]{Department of Astronomy, University of California, Berkeley, CA 94720-3411, USA}

\affil[11]{Instituto de Astrofísica, Facultad de Física and Centro de Astroingeniería, Pontificia Universidad Católica de Chile, Santiago, 7820436, RM, Chile}
            
\affil[12]{Millennium Institute of Astrophysics, Santiago, 7500011, RM, Chile}
            
\affil[13]{Space Science Institute, Boulder, 80301, CO, USA}

\affil[14]{Dipartimento di Fisica e Astronomia ``Augusto Righi", Università degli Studi di Bologna, Via Gobetti 93/2, Bologna, 40129, Italy}

\affil[15]{INAF -- Osservatorio di Astrofisica e Scienza dello Spazio di Bologna, Via Gobetti 93/3, Bologna, 40129, Italy}
            
\affil[16]{Department of Astronomy, University of Michigan, 1085 S University, Ann Arbor, MI 48109, USA}

\affil[17]{School of Astronomy and Space Science, Nanjing University, Nanjing, 210093, China}

\affil[18]{Key Laboratory of Modern Astronomy and Astrophysics (Nanjing University), Ministry of Education, Nanjing, 210093, China}

\affil[19]{Department of Astrophysics/IMAPP, Radboud University Nijmegen, P.O. Box 9010, Nijmegen, 6500 GL, The Netherlands}

\affil[20]{Department of Astronomy, Xiamen University, Xiamen, 361005, Fujian, China}

\affil[21]{Department of Astronomy, School of Physics, Huazhong University of Science and Technology, Wuhan, 1037, Hubei, China}

\affil[22]{Max-Planck-Institut für extraterrestrische Physik (MPE), Gie{\ss}enbachstra{\ss}e 1, Garching bei München, D-85748, Germany}

\affil[23]{Department of Physics, Anhui Normal University, Wuhu, Anhui 241002, China}

\affil[24]{Shanghai Astronomical Observatory, Chinese Academy of Sciences, Shanghai, 200030, China}

\affil[*]{Correspondence: xuey@ustc.edu.cn}
\affil[**]{Correspondence: lishuo@nao.cas.cn}

\begin{document}

\maketitle

\section*{GRAPHICAL ABSTRACT}

\includegraphics[width=\linewidth]{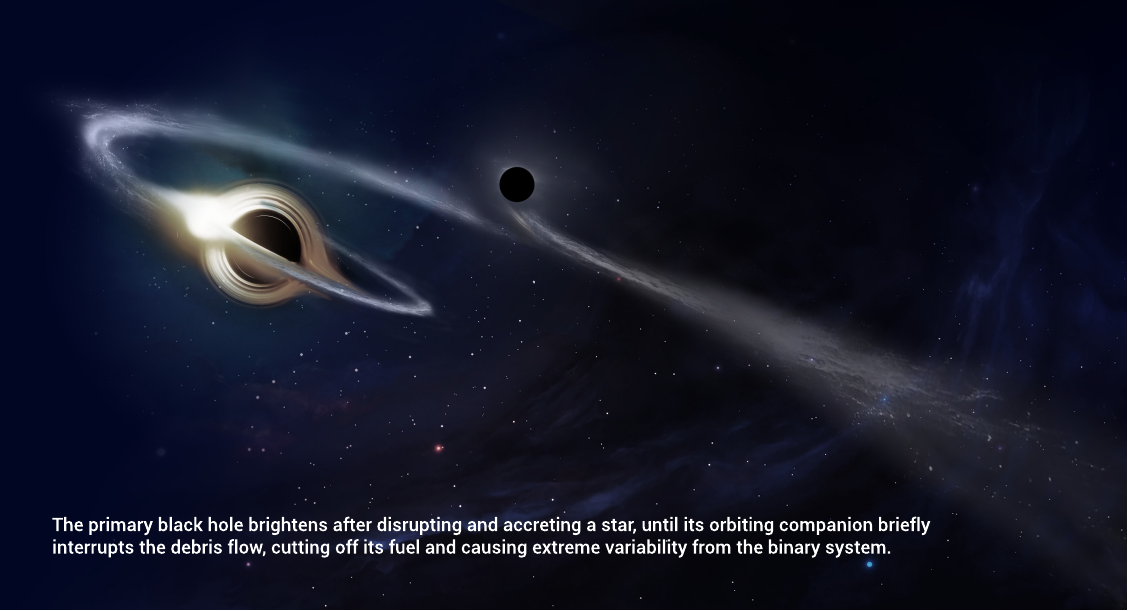}

\section*{PUBLIC SUMMARY}
\begin{itemize}

\item The source has the best X-ray coverage spanning $\sim 20$ years with the deepest total exposure of $\sim 10$ Ms.

\item The source is the faintest X-ray-selected tidal disruption event (TDE) to date.

\item The source displays unusually strong hard X-ray emission and extreme luminosity variability. 

\item The source is likely the most distant supermassive black hole binary TDE known to date.
\end{itemize}

\section*{ABSTRACT}

Tidal disruption events (TDEs), which occur when stars enter the tidal radii of supermassive black holes (SMBHs) and are subsequently torn apart by their tidal forces, represent intriguing phenomena that stimulate growing research interest and pose an increasing number of puzzles in the era of time-domain astronomy. Here we report an unusual X-ray transient, XID 935, discovered in the 7 Ms Chandra Deep Field-South, the deepest X-ray survey ever. XID 935 experienced an overall X-ray dimming by a factor of more than 40 between 1999 and 2016. Not monotonically decreasing during this period, its X-ray luminosity increased by a factor $> 27$ within 2 months, from $L_{\rm 0.5-7\ keV}<10^{40.87}$ erg s$^{-1}$ (10 October 2014 -- 4 January 2015) to $L_{\rm 0.5-7\ keV}=10^{42.31\pm 0.20}$ erg s$^{-1}$ (16 March 2015). The X-ray position of XID 935 is located at the center of its host galaxy with a spectroscopic redshift of 0.251, whose optical spectra do not display emission characteristics associated with an active galactic nucleus. The peak 0.5--2.0 keV flux is the faintest among all the X-ray-selected TDE candidates to date. Thanks to a total exposure of $\sim 9.5$ Ms in the X-ray bands, we manage to secure relatively well-sampled, 20-year-long X-ray light curves of this deepest X-ray-selected TDE candidate. We find that a partial TDE model could not explain the main declining trend. An SMBH binary TDE model is in acceptable accordance with the light curves of XID 935; however, it fails to match short-timescale fluctuations exactly. Therefore, the exceptional observational features of XID 935 provide a key benchmark for refining quantitative TDE models and simulations.

\section*{KEYWORDS}


Tidal Disruption Event, Supermassive Black Hole Binary, X-ray Transient

\section*{INTRODUCTION}

In the early 1960s, the first quasi-stellar object (i.e., quasar) was discovered.\cite{1963Natur.197.1040S} One possible power source of quasars was suggested \cite{hills_possible_1975} to be the tidal disruption of a star passing within the Roche limit of a supermassive black hole (SMBH), now known as a tidal disruption event (TDE). Later detailed research demonstrated that TDEs cannot release sufficient sustained energy to explain the persistent activity of an active galactic nucleus (AGN).\cite{young_black_1977, 1978MNRAS.184...87F} Nonetheless, an order of $10^{6}$ stars are sacrificed over the lifetime of a galaxy,\cite{1999MNRAS.309..447M} so disrupted stars may not contribute significantly to the growth of an SMBH with a mass larger than $10^{7}\ M_{\odot}$, TDEs may affect the spin distribution for BHs with mass $\sim 10^{6}\ M_{\odot}$ \cite{2019ApJ...877..143Z} and power observed AGNs in dwarf galaxies.\cite{2019MNRAS.483.1957Z} Although TDEs lost to galactic-scale gas \cite{2005SSRv..116..523F, 2015ARA&A..53..365N, 1999MNRAS.309..447M} in the race for providing fuel to AGNs, their unique characteristics are still an outstanding probe for revealing dormant SMBHs.\cite{rees_tidal_1988} Since NGC 5905 was first noticed,\cite{1999A&A...343..775K} over one hundred TDEs have been discovered, and probed with respectable amounts of multi-wavelength data. Recently, time-domain surveys \cite{sazonov_first_2021, 2023ApJ...942....9H} have propelled TDE research into a new era of proactive exploration, rather than relying on serendipitous discoveries in legacy data.

The TDE model described in Rees (1988) \cite{rees_tidal_1988} neglects partial stellar disruptions, now known as the partial TDE (pTDE), in which the pericenter of the star is slightly farther from the tidal radius, resulting in a surviving stellar core after the encounter.\cite{2013ApJ...767...25G} In the case of pTDE, the power law of the fallback rate is predicted to be steeper, $\propto t^{-9/4}$,\cite{2019ApJ...883L..17C} compared to complete disruption, $\propto t^{-5/3}$.\cite{phinney_manifestations_1989} Furthermore, a remnant core has the potential to produce a second TDE if it becomes bound to the SMBH after the first encounter.\cite{2020ApJ...904..100R}

A regular TDE usually refers to the stellar disruption by a single quiescent SMBH lying at the center of a galaxy.\cite{rees_tidal_1988, phinney_manifestations_1989, 2011Natur.476..421B} 
However, special TDEs occurring in SMBH binaries (SMBHBs) have been proposed,\cite{liu_interruption_2009} which exhibit interrupted tidal flares caused by the companion black hole, compared to the power-law decay light curve (i.e., $\propto t^{-5/3}$).\cite{phinney_manifestations_1989} SDSS J120136.02$+$300305.5 (hereafter SDSS J1201$+$30),\cite{liu_milliparsec_2014} whose deep dips in its $\sim 300$-day evolving X-ray light curve could be explained by the SMBHB TDE model, is the first SMBHB TDE candidate with a spectroscopic redshift of $z_{\rm spec}=0.146$. Recently, a second X-ray candidate at $z_{\rm spec}=0.1655$ with a $\sim 550-$day X-ray light curve, OGLE16aaa, was reported.\cite{shu_x-ray_2020} 
Both SDSS J1201$+$30 and OGLE16aaa exhibit apparent interruptions in their soft X-ray light curves, which can be explained by the companion black hole perturbing the fallback rate of disrupted debris. The 0.2--2.0 keV X-ray flux of SDSS J1201$+$30 decreased by more than 47 times within 7 days, while the 0.3--2.0 keV X-ray flux of OGLE16aaa increased by more than 63 times within 9 days. No intense hard X-ray emission was observed in either source.

To date, most X-ray TDEs have been discovered through the detection of soft X-ray flares, produced by emission from the transient accretion disk formed after the circularization of stellar debris. Near peak luminosities, the X-ray spectra of these TDEs are well fitted by either a blackbody with a temperature of $kT_{\rm bb}=0.04-0.12$~keV or a power law with an index of $\Gamma_{X}=4-5$, followed by a spectral hardening over time.\cite{2020SSRv..216...85S} In addition, there is another rare population of X-ray TDEs characterized by much harder spectra, extremely high X-ray luminosities, and dramatic short-time variability.\citep{2011Natur.476..421B,2022Natur.612..430A} They are believed to be produced by powerful relativistic jets aligned with our line of sight, rather than by accretion.

Here, we report an unusual X-ray transient, XID 935 ($\mathrm{RA}_\mathrm{J2000}=53.248664^{\circ}$, $\mathrm{Dec}_\mathrm{J2000}=-27.841828^{\circ}$, with a 1-$\sigma$ positional accuracy of 0.3 arcsec), discovered in the deepest X-ray survey ever---the 7 Ms Chandra Deep Field-South (CDF-S),\cite{luo_chandra_2016, xue_chandra_2017} whose extraordinary observational properties make it the deepest X-ray-selected TDE candidate. 
We use a cosmology of $H_{0} = 70\ \rm{km\ s^{-1}\ Mpc^{-1}}$, $\Omega_{M} = 0.27$, and $\Omega_{\Lambda} = 0.73$.

\section*{MATERIALS AND METHODS}

\subsection*{X-ray data}
\label{X-ray Data}

\subsubsection*{Data reduction}

The CDF-S\cite{luo_chandra_2016} and Extended-CDF-S (E-CDF-S; 250-ks depth)\cite{2005ApJS..161...21L,xue_2_2016} surveys include a total of 111 Chandra observations performed between 1999 and 2016. The XMM deep survey in the CDF-S (XMM-CDFS; 3-Ms depth)\cite{ranalli_xmm_2013} consists of 33 observations carried out between 2001--2002 and 2008--2010. The Wide-CDF-S (W-CDF-S; 29-ks depth)\cite{2021ApJS..256...21N} has one observation covering the position of XID 935. Table S1 lists the observation identification number (ObsID), date, exposure, and bin name of these X-ray observations. Observation 581 is not included in the CDF-S due to telemetry saturation, but XID 935 is located on CCD I0 which escaped this disaster. All CDF-S and XMM-CDFS observations cover the position of XID 935. The E-CDF-S consists of four distinct sub-fields with only the edges overlapping between them, so only 3 (ObsIDs = 5021, 5022, and 6164) out of its 9 observations observed XID 935 and are included in this work. ROSAT\cite{1982AdSpR...2d.241T} and Swift\cite{2005SSRv..120..165B} observations are listed in Table S2.

CIAO 4.14 with CALDB 4.9.7 was utilized to process the cleaned event files from the 7 Ms CDF-S \cite{luo_chandra_2016} and E-CDF-S.\cite{xue_2_2016} The \emph{dmellipse} of CIAO was used to generate an elliptical source region that encloses 90\% of the flux. The background region is an annulus with an inner radius of $R_{\rm in} =12^{\prime \prime}$ and an outer radius of $R_{\rm out} =40^{\prime \prime}$. According to the 7 Ms CDF-S catalog, six sources were masked since they are immediately adjacent to the background region of XID 935, as shown in Figure S2. In certain observations, the background region overlapped with the CCD gap, and these areas were masked using box regions. Spectral files were generated by the CIAO tool \emph{specextract} and combined using \emph{combine\_spectra} according to the time bins specified in Table S1. The parameter \emph{correctpsf} in \emph{specextract} was set to yes for aperture correction. 

SAS 21.0.0 was utilized to obtain cleaned event files of XMM-CDFS observations following the threads of {\it sas-thread-startup}, {\it sas-thread-epic-reprocessing}, and {\it sas-thread-epic-filterbackground}. Subsequently, the spectra were extracted using the spectral analysis threads of MOS ({\it sas-thread-mos-spectrum}) and PN ({\it sas-thread-pn-spectrum}). The source region is a circle with a radius of $10^{\prime \prime}$ (corresponding to 60\% encircled energy at 1.5 keV), and the background region is an annulus with an inner radius of $R_{\rm in} =25^{\prime \prime}$ and an outer radius of $R_{\rm out} =60^{\prime \prime}$. According to the 7 Ms CDF-S catalog, nine sources were masked with a circle with a radius of $20^{\prime \prime}$, since they overlap with the background region of XID 935, as shown in Figure S3.
The {\it arfgen} applied default aperture corrections for point sources. The spectral combination was performed by the SAS tool \emph{epicspeccombine} according to the time bins listed in Table S1.

\emph{Xselect} of HEASoft was used to extract the spectra from the event files of ROSAT and Swift (see Table S2 for ObsIDs). XID 935 is not detected in the ROSAT All-Sky Survey (RASS) observations conducted between June 1990 and August 1991. For Swift, the source region is a circle with a radius of $R =30^{\prime \prime}$ and the background region is an annulus with an inner radius of $R_{\rm in} =40^{\prime \prime}$ and an outer radius of $R_{\rm out} =80^{\prime \prime}$. For ROSAT, the region configuration is the same as that of Swift, but $R =60^{\prime \prime}$, $R_{\rm in} =70^{\prime \prime}$ and $R_{\rm out} =120^{\prime \prime}$. The arf file was generated using \emph{xrtmkarf} of HEASoft for Swift, with a setting parameter $\emph{psfflag}=yes$ to correct the point spread function for point-like sources. The \emph{pcarf} of HEASoft was used to generate the arf file for ROSAT, with an aperture-correction factor of 0.94 based on the encircled energy function. For eROSITA,\cite{2024A&A...682A..34M} the eROSITA-DE Data Release 1 archive supports a tool for the upper limit for a single position. 

\subsubsection*{Bin strategies}

The significance of source detection is calculated using the likelihood-ratio test (LRT),\cite{2008NIMPA.595..480C} see Listing S1. For p-values larger than 0.0027, the flux/luminosity is considered as the upper limit at the $90\%$ confidence level. For p-values below this threshold  (i.e., $\geq$ 3 $\sigma$), the flux/luminosity is reported as the mean value with a 1-$\sigma$ uncertainty. Two different binning strategies for the light curve are applied to the Chandra and XMM-Newton data:

\noindent
A. 1-month binning: 

(1) For both Chandra and XMM-Newton observations, a 1-month-bin scheme is initially applied for a straightforward overall view of the light curves (see Figure S4 for a flow diagram and Figure S5 for the 1-month-bin light curves).

\noindent
B. adaptive binning:

(1) Observations are marked as detected points if the significance of likelihood-ratio test in the full band is greater than 3 $\sigma$, and as non-detected points if the significance is less than 3 $\sigma$.

(2) For non-detected points, a 1-month-bin scheme is applied, but it stops at any detected point (see Figure S6 for a flow diagram). New bins are marked as non-detected points if the significances of likelihood-ratio tests in the full, soft, and hard band are all less than 3 $\sigma$; otherwise, they are marked as detected points.

(3) For detected points (both from step 1 and step 2), a 1-month-bin scheme is applied, but it stops at any non-detected point. If the right margin of a bin coincides with the exposure duration of a point, the right margin is set as the start time of that point.

(4) Some data remain non-detection even after one-month binning. For these non-detected points, a 4-months-bin scheme is applied, but it stops at any detected point. New bins are marked as non-detected points if the significances of likelihood-ratio tests in the full, soft, and hard band are all less than 3 $\sigma$; otherwise, they are marked as detected points.

\subsubsection*{Spectral fitting}

\begin{figure}[htbp]
    \centering
    \includegraphics[width=\linewidth]{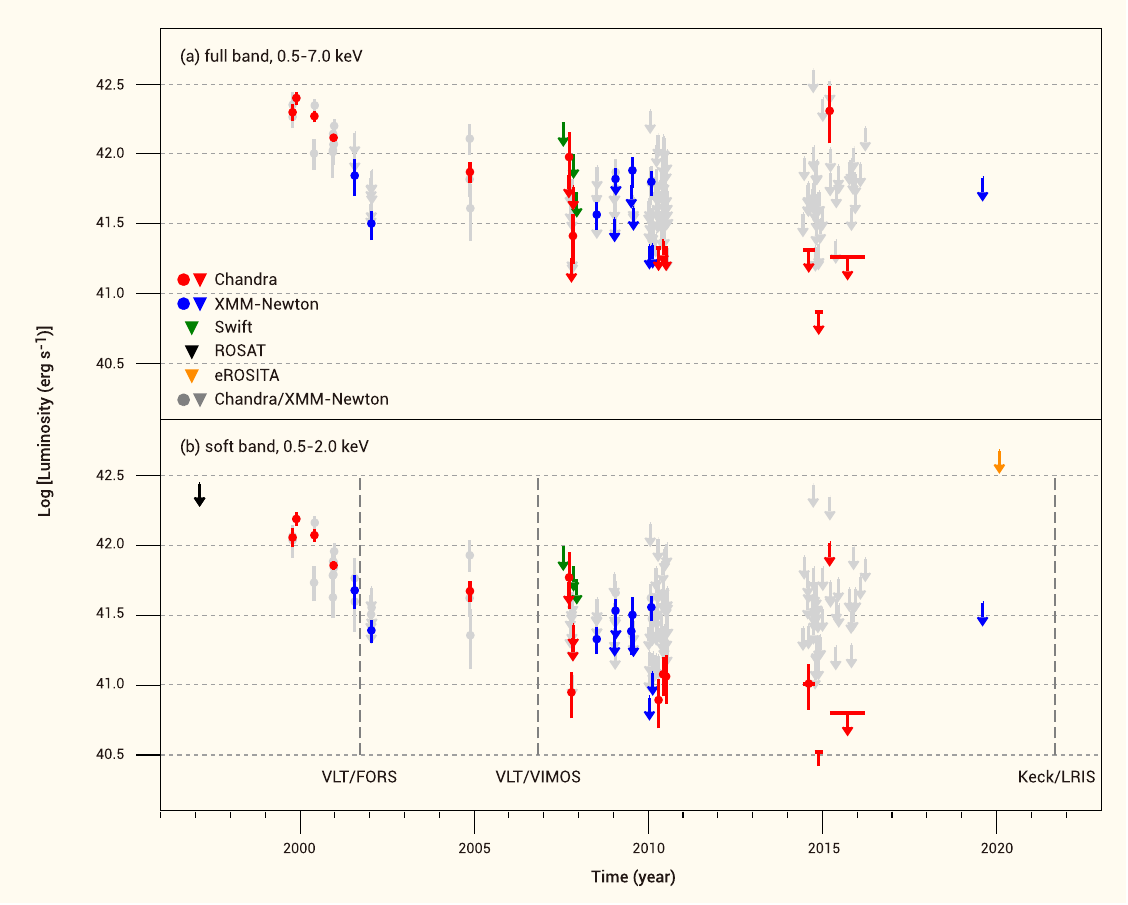}
    \caption{\textbf{Adaptive-bin X-ray light curves of XID 935.} (a) Light curve in the rest-frame 0.5--7.0 keV band (full band, containing $768.9\pm43.3$ and $905.0\pm84.0$ Chandra and XMM-Newton net counts, respectively).
    (b) Light curve in the rest-frame 0.5--2.0 keV band (soft band, containing $595.9\pm28.5$ and $680.7\pm62.6$ Chandra and XMM-Newton net counts, respectively). 
    Red, dark blue, green, black, and orange data are from Chandra, XMM-Newton, Swift, ROSAT, and eROSITA, respectively. 
    Grey data represent individual observations of Chandra and XMM-Newton.
    Horizontal bars indicate ranges of time bins. Arrows indicate 90\%-confidence-level upper limits.
    The data gaps between 2002–2004 and 2005–2007 indicate no X-ray observational coverage.
    Dashed lines denote the observation times of optical spectra.}
    \label{bin_adaptive}
\end{figure}

Given that XID 935 has low counts in most observations, it is not suitable to fit those spectra. The \emph{aprates} of CIAO is used to calculate the count rate, then \emph{modelflux} of CIAO is used to calculate the flux using this count rate, and subsequently the luminosity is derived from this flux. For \emph{modelflux}, the parameters $model=``xspowerlaw.pow1"$, $absmodel=``xsphabs.abs1"$, $absparams=``abs1.nH=0.008"$ \cite{1992ApJS...79...77S} and $paramvals=``pow1.PhoIndex=2.4"$ are based on the spectral fitting result through stacking all XID 935 Chandra observations. The cosmology model is set as the default one in XSPEC. Figure \ref{bin_adaptive} shows the adaptive-bin scheme light curves; data listed in Table S4.
Spectral fitting is only performed for the spectrum with net counts larger than 60 (see Table S3, and Figure S7). To fit the spectrum for a weak source, especially with low counts in its background, a better grouping strategy is to ensure that each bin in the background spectrum contains at least one count to use the W statistic.\cite{bias_in_profile_Poisson_likelihood} The HEASoft tool \emph{ftgrouppha} supports this grouping by setting the parameter $grouptype=bmin$ and $groupscale=1$. The XSPEC 12.13.0 software was used to fit the regrouped spectra derived from both Chandra and XMM-Newton observations. We apply a Galactic and intrinsic absorbed redshifted power-law model, $tbabs\times ztbabs \times zpowerlw$, to the spectra of XID 935, with the Galactic column density fixed at $N_{\rm H} = 8.8 \times 10^{19}\ {\rm cm}^{-2}$, redshifts of $zpowerlw$ and $ztbabs$ fixed at $z=0.251$, power-law photon index and redshifted $N_{\rm H}$ left free to vary. The best-fit column density during these bins was always $N_{\rm H} < 10^{22}\ {\rm cm}^{-2}$ (see Section \nameref{Exclusion of Variable Absorption}), which implies that the absorption is not severe in these periods.

\subsection*{UV/optical data}
\label{UVOdata}

The UV/optical observations of XMM/OM and Swift/UVOT were performed for XID 935. The HEASoft tool \emph{uvotimsum} was utilized to sum the Swift/UVOT UV/optical observations and the SAS tool \emph{ommosaic} was used to sum the XMM/OM UV/optical observations. Photutils \cite{larry_bradley_2023_7946442} was used to perform aperture photometry (a circle with a $4^{\prime \prime}$ radius for the source region; an annulus with a $6^{\prime \prime}$ inner radius and a $10^{\prime \prime}$ outer radius for the background region) on the summed UV/optical images. The aperture correction factors are 0.65, 0.7, 0.7, and 0.8 for XMM/OM UVW2, UVM2, UVW1, and u bands, respectively. For Swift/UVOT UVW2, UVM2, UVW1, and u bands, the aperture correction factor is all 0.95. The results are shown in Figure S1.

\subsection*{Host SED fitting}
\label{SED Fitting}

\begin{figure}[htbp]
    \centering
	\includegraphics[width=0.7\columnwidth]{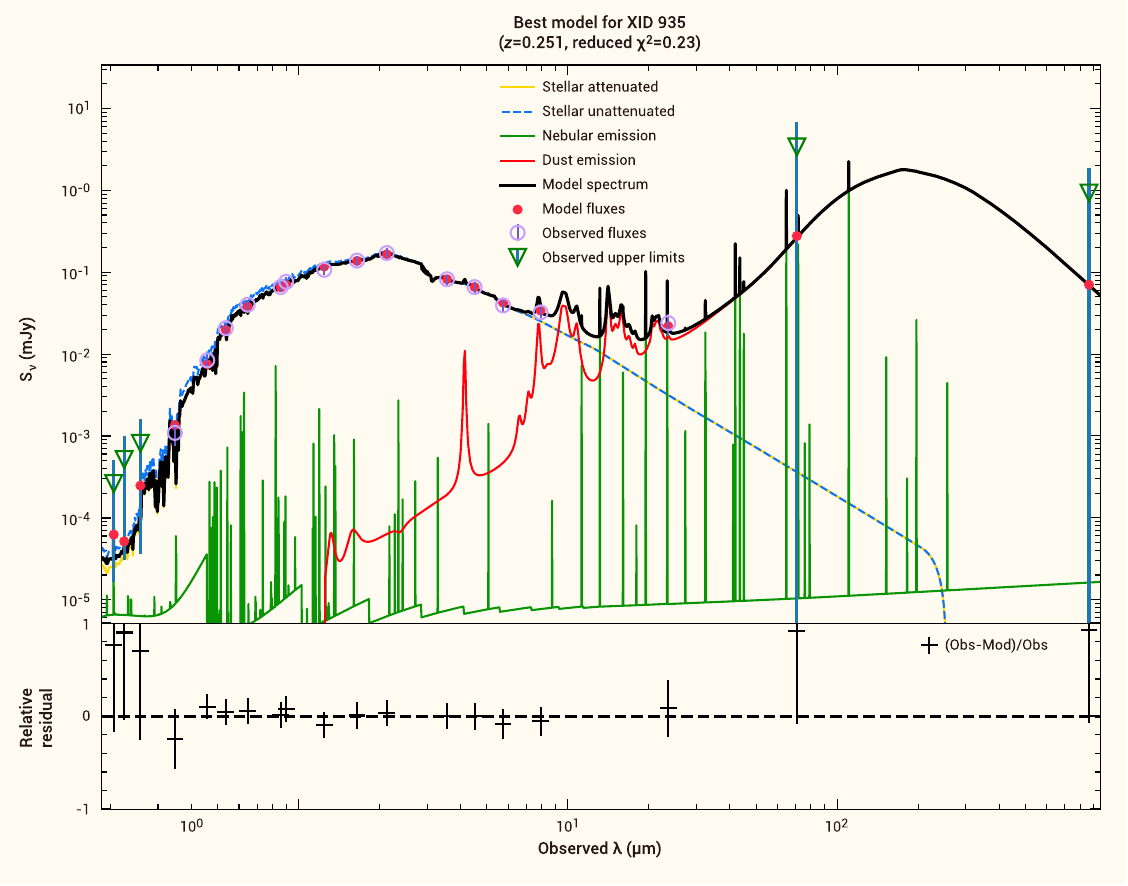}
    \caption{\textbf{SED fitting result for XID 935 using CIGALE.} The stellar mass $M_{\star}=(3.1\pm0.6)\times10^{10}\ M_{\odot}$ and the star-formation rate SFR$<1.3\times10^{-2} M_{\odot}$/year\ (90\%\ confidence level) were obtained. Note that an AGN component is not required in the fitting.}
    \label{SEDbestmodle}
\end{figure}

The SED fitting was performed using CIGALE \cite{yang_fitting_2022} with the following multi-wavelength data: 
$WFI\_B$, $WFI\_V$, $WFI\_R$, $WFI\_I$, $WFI\_z$, $WFI\_J$, $WFI\_H$, $WFI\_K$, $spitzer.irac.ch1$ ($3.6\ \mu m$), $spitzer.irac.ch2$ ($4.5\ \mu m$), $spitzer.irac.ch3$ ($5.8\ \mu m$), $spitzer.irac.ch4$ ($8.0\ \mu m$);\cite{damen_simple_2011} $spitzer.mips.24$ ($24\ \mu m$);\cite{elbaz_goods_2011} $spitzer.mips.70$ ($70\ \mu m$);\cite{magnelli_evolution_2011} $scuba.850$ ($850\ \mu m$);\cite{barger_submillimeter_2019} and $Swift\_$ $UVOT.UVW2$, $Swift\_$$UVOT.UVM2$, $Swift\_$$UVOT.UVW1$, $Swift\_$$UVOT.U$ bands obtained as described in Section \nameref{UVOdata}. For each band, based on $A_{f}/A_{v}$ ((SVO Filter Profile Service:\href{http://svo2.cab.inta-csic.es/theory/fps/}{http://svo2.cab.inta-csic.es/theory/fps/}) 
and $A_{v} = R_{v} * E(B-V)$, setting $E(B-V)=0.0072$, assuming $R_{v}=3.1$, the $A_{v}$ has been calculated and applied in Galactic extinction correction. Table S6 lists the CIGALE parameters used in our fitting and Figure \ref{SEDbestmodle} shows the fitting result.
We use GALFIT \cite{2002AJ....124..266P} to decompose the bulge and disk components of images in the $WFI\_B$, $WFI\_V$, $WFI\_I$, and $WFI\_z$ bands. SED fitting for the bulge reveals no significant difference from the total galaxy.

\subsection*{Optical spectral reduction}
\label{optical spectrum reduction}

\begin{figure}[htbp]
    \centering
    \includegraphics[width=0.7\linewidth]{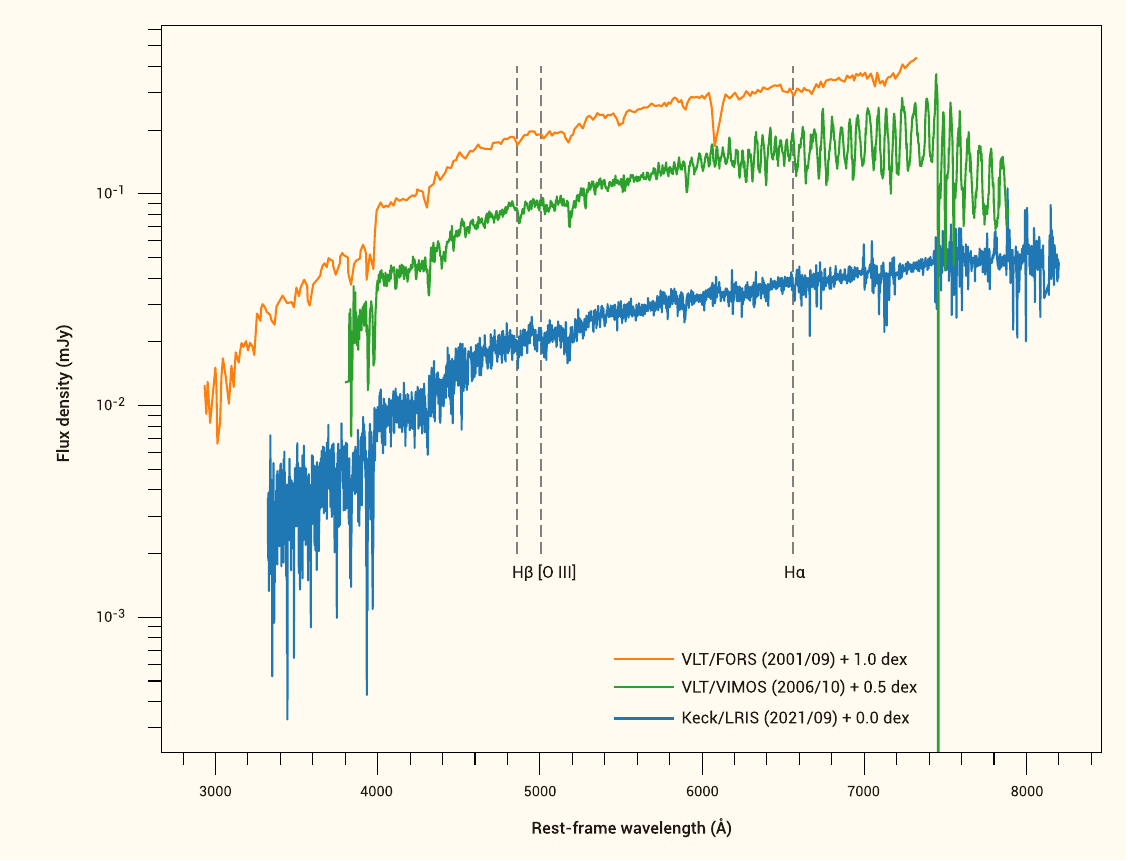}
    \caption{\textbf{Optical spectra of XID 935.} The spectral analysis of the Keck/LRIS spectrum (blue), performed on 8 September 2021, shows that the host galaxy of XID 935 has no significant AGN characteristics. The flux upper limit of the [O \Rmnum{3}] emission line is $10^{-16.68}$ erg s$^{-1}$ cm$^{-2}$ and the stellar velocity dispersion is $138\pm{31}$\ km/s. The VLT/VIMOS spectrum (green) was obtained on 30 October 2006 (Balestra et al. 2010). The VLT/FORS spectrum (orange) was observed on 18-20 September 2001 (Szokoly et al. 2004).}
    \label{opspec.pdf}
\end{figure}

To pin down the nature of XID 935 whose temporal evolution could be due to the AGN variability, we obtained an optical spectrum of the source with the Keck I telescope at the W. M. Keck Observatory and the Low Resolution Imaging Spectrometer (LRIS; \citealp{1995PASP..107..375O}) on UT2021-09-08 14:30 (MJD 59465.6045). A total of 850 s and 450 s $\times$2 exposures were obtained through the LRIS blue- and the red-arm CCDs, respectively. The combination of the 5600 \AA\ dichroic, the 600/4000 blue grism (with a 0.63\,\AA\,pixel$^{-1}$ dispersion) and the 400/8500 red grism (with a 1.20\,\AA\,pixel$^{-1}$ dispersion), together with a 1$''$ slit width provided an observed wavelength coverage of 3150--10270\,\AA. The spectral resolution near the central wavelength of the blue and the red arms yields, R$_{\rm blue}\sim$942 and R$_{\rm red}\sim$844, corresponding to $\sim$4.7\,\AA\ and $\sim$9.4\,\AA\ resolution element in the blue and the red arms, respectively. During the observation, the slit was aligned to the parallactic angle, which is $\approx-5$ degrees from the North. The raw LRIS spectral data were processed by the fully automated LPipe pipeline.\citep{2019PASP..131h4503P} The spectra were dominated by absorption lines without prominent emission lines, indicating the absence of significant AGN activity. To assess the upper limits of AGN activity, we calculated the upper limits of [OIII] as follows.  The Keck/LRIS spectrum was first corrected for Galactic extinction using the dust map of Schlafly et al. (2011)\cite{2011ApJ...737..103S} and the extinction curve from Fitzpatrick et al. (1999).\cite{1999PASP..111...63F} The starlight continuum was then modeled using the Python package pPXF of Cappellari et al. (2004)\cite{2004PASP..116..138C} and Cappellari et al. (2017),\cite{2017MNRAS.466..798C} which enables the extraction of stellar population and kinematics through penalized fitting. Figure \ref{opspec.pdf} shows the optical spectra.
After subtracting the modeled continuum from the spectrum, we calculated the [OIII] upper limits at a 90\% confidence level using the method outlined in Avni et al. (1976),\cite{1976ApJ...210..642A} assuming a pseudo-width of [OIII] from composite spectra of Vanden et al. (2001).\cite{2001AJ....122..549V}

\subsection*{Numerical modeling}
\label{subsec_fit}

To verify whether the SMBHB TDE model can explain the observed results, we conducted a series of numerical simulations for comparison with the observational data. The numerical model is based on three-body scattering experiments. Here, we briefly introduce our modeling approach. For more details, please refer to Liu et al. (2009) and Liu et al. (2014).\cite{liu_interruption_2009,liu_milliparsec_2014}

We assume that a star on a parabolic orbit is tidally disrupted at the pericenter $r_{\rm p}$. The specific energy $E$ of the stellar debris after disruption ranges from $-E_{\rm b}-\Delta E$ to $-E_{\rm b}+\Delta E$, where $E_{\rm b}$ is the orbital binding energy of the star, $\Delta E$ is the spread in specific energy.\cite{rees_tidal_1988, 1989ApJ...346L..13E} For a star on the parabolic orbit, nearly half of the debris is bound to the SMBH after the tidal disruption. As a result, the specific energy of the bound debris ranges from $-\Delta E$ to 0. We assume that the disrupted star is constructed as polytropes with a solar mass. And the mass distribution of the debris can be assumed to be constant.\cite{1989ApJ...346L..13E} To simplify the calculation, we neglect the interactions between debris elements. As a consequence, each element follows a ballistic trajectory. Thus we can perform scattering experiments for each three-body system composed of the binary black holes and the debris element. Although this simplification is less precise than full hydrodynamical simulations, it provides a sufficiently accurate approximation for order-of-magnitude estimates and qualitative analysis.

From the scattering experiment results, it is easy to estimate the rate at which the debris returns to the vicinity of the black hole, namely the mass fallback rate. Assuming that the fallback rate is proportional to the X-ray emission, we can make the conversion factor $\eta$ as a free parameter and fit it to the observed flux. By comparing the “light curve” derived from the scattering experiments (more precisely, the fallback curve) with the observational data, we can infer the approximate time of disruption based on the position of the peak along the time axis. This gives us $T_{\rm delay}$, the delay between the actual disruption and the first effective observational detection.

It should be noted that the methodology described above may lead to multiple sets of parameters that fit the data. Our analysis serves to qualitatively evaluate the plausibility of the SMBHB TDE model in explaining the observations, rather than to exactly determine the system’s actual orbital parameters.

\section*{RESULTS AND DISCUSSIONS}

\subsection*{Light curves and spectra}
\label{Light Curves and Spectra}

XID 935 was originally detected in the first 1 Ms observations of the CDF-S,\cite{2002ApJS..139..369G} with an optical counterpart subsequently classified as an inactive galaxy showing only absorption lines with $z_{\rm spec}=0.251$ based on a VLT/FORS optical spectrum obtained on 18--20 September 2001 \cite{szokoly_chandra_2004} as shown in Figure \ref{opspec.pdf}. Luo et al. (2017) \cite{luo_chandra_2016} classified XID 935 as an AGN based on the entire 7 Ms CDF-S data. 
Zheng et al. (2017)\cite{zheng_deepest_2017} found six candidate transients based on the criterion that the flux change exceeds 3-$\sigma$ of the average flux, using the 3-month-bin 7 Ms CDF-S light curves. Three of these candidates (XID 725,\cite{2017MNRAS.467.4841B} XID 330,\cite{2019Natur.568..198X} and XID 403 \cite{2023ApJ...949....6Y}) have been reported in the literature. XID 935 is among these six candidate transients and is reclassified as a long outburst transient. XID 935 was additionally observed as part of the E-CDF-S (250-ks depth),\cite{2005ApJS..161...21L,xue_2_2016} the XMM-CDFS (3-Ms depth),\cite{ranalli_xmm_2013} the W-CDF-S (29-ks depth),\cite{2021ApJS..256...21N} Swift (0.5-Ms depth),\cite{2005SSRv..120..165B} ROSAT (8.52-ks depth),\cite{1982AdSpR...2d.241T} and eROSITA (0.218-ks depth),\cite{2024A&A...682A..34M} providing relatively well-sampled 20-year-long X-ray light curves of XID 935. Unlike the 3-month bin scheme adopted by Zheng et al. (2017),\cite{zheng_deepest_2017} we use an adaptive-bin scheme to obtain light curves (see Figure \ref{bin_adaptive}; data listed in Table S4), given that some interrupted flares that are shorter than 3-month time scale will be smoothed out and that stacking adjacent non-detection observations could increase the detection sensitivity.

Below we summarize the observational features of XID 935: 

(1) A peak soft X-ray luminosity of $L_{\rm 0.5-2\ keV}=10^{42.14\pm 0.04}$ erg s$^{-1}$ was recorded in October-November 1999, dimming by a factor of at least 40 between November 1999 and October 2014; 

(2) Significant hard X-ray emission of $L_{\rm 2-7\ keV}=10^{42.09\pm 0.12}$ erg s$^{-1}$ was detected in November 1999; however, there was almost no detection ($< 3 \sigma$) of hard X-ray emission after December 2000 except on March 16, 2015;

(3) It displays extreme and fast luminosity variations: the full-band (0.5--7.0 keV) X-ray luminosity increased by a factor of at least 27 within $\sim 2$ months, from a $90\%$-confidence-level upper limit of $L_{\rm 0.5-7\ keV}<10^{40.87}$ erg s$^{-1}$ (10 October 2014 -- 4 January 2015) to $L_{\rm 0.5-7\ keV}=10^{42.31\pm 0.20}$ erg s$^{-1}$ (16 March 2015); 

(4) The X-ray emission is centered in the host galaxy, with no AGN characteristics (see Figure \ref{opspec.pdf}) or UV/optical variability (see Figure S1);

(5) No radio emission was detected at 15 mm,\cite{2014MNRAS.439.1212F} 6 cm,\cite{2012MNRAS.426.2342H} 13 cm,\cite{2012A&A...544A..38Z} 21 cm,\cite{2008ApJS..179..114M,2013ApJS..205...13M,2006AJ....132.2409N,2014MNRAS.440.3113H,1998AJ....115.1693C} 92 cm,\cite{2007ASPC..380..243A} 1.25--3.75 m,\cite{2015PASA...32...25W} 1.5--3 m,\cite{2014PASA...31...45H} and 4 m bands,\cite{2007AJ....134.1245C} as well as the Very Large Array (VLA) Survey of the CDF-S,\cite{2008ApJS..179...71K} the VLA Survey of the GOODS-S,\cite{2019ApJ...875...80G} the ultradeep radio imaging of GOODS-S/HUDF,\cite{2020ApJ...901..168A} and the VLA Sky Survey (VLASS),\cite{2020PASP..132c5001L} with no jet being detected.

\subsection*{Exclusion of a variable absorption origin}
\label{Exclusion of Variable Absorption}

The X-ray variability could be attributed to intrinsic spectral variation and/or variable absorption. For an absorbed power-law model, with a power-law photon index of $\Gamma = 2.4$ (derived from spectral stacking of Chandra observations listed in Table S1) and a column density of $N_{\rm H} = 1 \times 10^{22}\ {\rm cm}^{-2}$, a minimum of 60 full-band net counts are required for a reliable fit. However, since most of the bins for XID 935 have low counts, only seven bins contain full-band net counts exceeding 60, which are used to estimate the intrinsic absorption. A Galactic and intrinsic absorbed redshifted power-law model, $tbabs\times ztbabs \times zpowerlw$, is applied to these seven spectra, with the Galactic column density fixed at $N_{\rm H} = 8.8 \times 10^{19}\ {\rm cm}^{-2}$ \cite{1992ApJS...79...77S} and the redshift fixed at $z=0.251$.\cite{szokoly_chandra_2004} The power-law photon index and the redshifted $N_{\rm H}$ are allowed to vary freely. The best-fit column density in these seven bins (see Table S3) was always less than $10^{22}\ {\rm cm}^{-2}$. If the X-ray variability were primarily driven by varying obscuration, a harder spectrum would be expected, as soft photons would be absorbed. However, after 2001, most bins of XID 935 show no detection in the hard band but significant detection in the soft band, suggesting a preference for soft spectra. When both the soft and hard bands show non-detection, it becomes challenging to disentangle the contribution from intrinsic spectral variation and variable absorption. Notably, the optical spectra of XID 935 show no characteristics of AGN activity, implying that there is little surrounding matter near the black hole. Hence, we assume that the X-ray variability of XID 935 is not caused by the variable absorption, and a Galactic-absorbed redshifted power-law model, $tbabs\times zpowerlw$, with fixed $N_{\rm H} = 8.8 \times 10^{19}\ {\rm cm}^{-2}$, $\Gamma = 2.4$, and $z=0.251$, is adopted to estimate the flux and luminosity (see materials and methods).

\subsection*{Exclusion of an AGN origin}

It is not uncommon for AGNs to vary in brightness by several dozen times, either due to intrinsic spectral variation or variable absorption. Such sources typically exhibit persistent AGN signatures revealed by the strong narrow emission lines in their optical spectra. However, optical spectra of XID 935 obtained with VLT/FORS in September 2001 \cite{szokoly_chandra_2004} and VLT/VIMOS in October 2006 \cite{2010A&A...512A..12B} reveal no apparent AGN signature. We utilized Keck/LRIS to obtain a new optical spectrum of XID 935 on 8 September 2021 (see Figure \ref{opspec.pdf} and materials and methods), which also shows no AGN signature. 
From the Keck spectrum, we estimate the 90\%-confidence-level flux upper limit of the [O \Rmnum{3}] emission line, which is $f_{\rm [O \Rmnum{3}]} < 10^{-16.68}$ erg s$^{-1}$ cm$^{-2}$. In addition, the ZTF r-band light curve shows no statistically significant variability over timescales of 5 years (see Figure S1). These findings suggest that the host galaxy of XID 935 is likely not an active galaxy.
Some AGNs, known as ``X-ray bright, optically normal galaxies" (XBONGs),\cite{2009ApJ...706..797T} are X-ray bright but lack apparent AGN optical signatures. However, to our knowledge, no XBONG has shown X-ray variability by a factor of $\geq 40$. Moreover, the AGN scenario fails to account for the rapid luminosity changes discussed in point (3) of Section \nameref{Light Curves and Spectra}, so we exclude this possibility from further consideration.

\subsection*{Exclusion of an X-ray binary origin}

X-ray binary (XRB) populations could also be a possible origin of hard X-ray emission. Using the stellar mass [$M_{\star}=(3.1\pm0.6)\times10^{10}\ M_{\odot}$] and the star formation rate (SFR$<1.3\times10^{-2} M_{\odot}$ yr$^{-1}$, at the 90\% confidence level) derived from spectral energy distribution (SED) fitting (see materials and methods), and applying the relationship between XRB X-ray luminosity, stellar mass, SFR, and redshift,\cite{lehmer_evolution_2016} we estimate the contribution of XRB populations to the X-ray emission of XID 935 to be $L_{2-10\ \mathrm{keV}}(\rm XRB)=10^{40.22\pm0.30}$ erg s$^{-1}$. Apparently, this contribution is insufficient to account for the observed hard X-ray emission ($L_{2-7\ \mathrm{keV}}\sim 10^{42}$ erg s$^{-1}$). Moreover, due to the large number of XRBs, these populations as a whole are not expected to exhibit significant variability. 
Essentially no known individual extragalactic ultraluminous X-ray sources (ULXs) exhibit the required luminosity of $L_{2-7\ \mathrm{keV}}\sim 10^{42}$ erg s$^{-1}$, as ESO 243-49 HLX-1,\cite{2009Natur.460...73F} one of the most luminous ULXs known, attains only $L_{2-7\ \mathrm{keV}}\sim 10^{41}$ erg s$^{-1}$. Given that the number of galaxies in the filed of view of 7 Ms CDF-S\cite{luo_chandra_2016} reaches $50539$, the probability of a Galactic or local universe XRB aligning randomly with the core of a galaxy (within a $0.3^{\prime \prime}$ radius) is about 0.79\%. Therefore, XRB is not a reasonable explanation for XID 935.

\subsection*{TDE scenarios}

\begin{figure}[htbp]
    \centering
    \includegraphics[width=\linewidth]{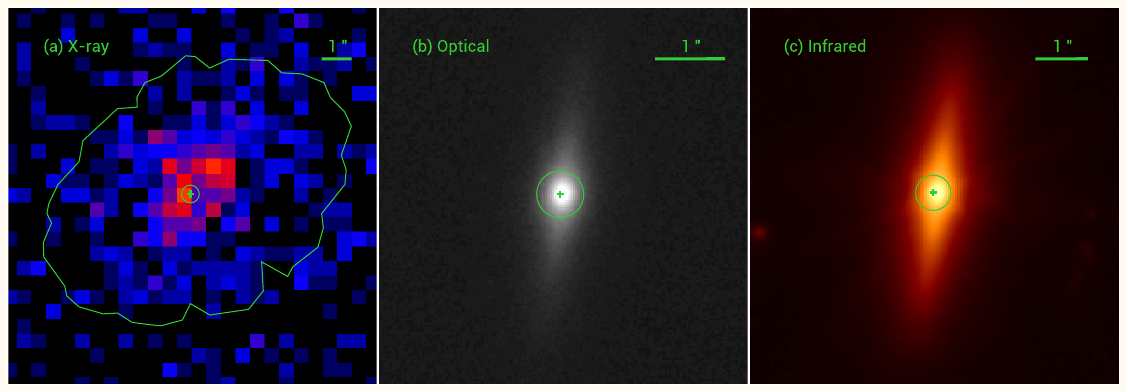}
    \caption{\textbf{Multi-wavelength images of XID 935.} (a) 7 Ms CDF-S image in the observed 0.5--7.0 keV band, with total net counts of $768.9\pm43.3$. The green polygon is the 90\% encircled-energy fraction region, centered at the source position (green plus; with a 1-$\sigma$ positional accuracy of 0.3 arcsec, indicated by the green circle, corresponding to 0.99 kpc at $z=0.251$).  (b) GOODS z band image. (c) JWST-NIRCam F227W band image. The green pluses and circles in Panels (b) and (c) are the same as that in Panel (a). The position of XID 935 is consistent with the nucleus of its host galaxy.}
    \label{host.pdf}
\end{figure}

\begin{figure}[htbp]
    \centering
    \includegraphics[width=\linewidth]{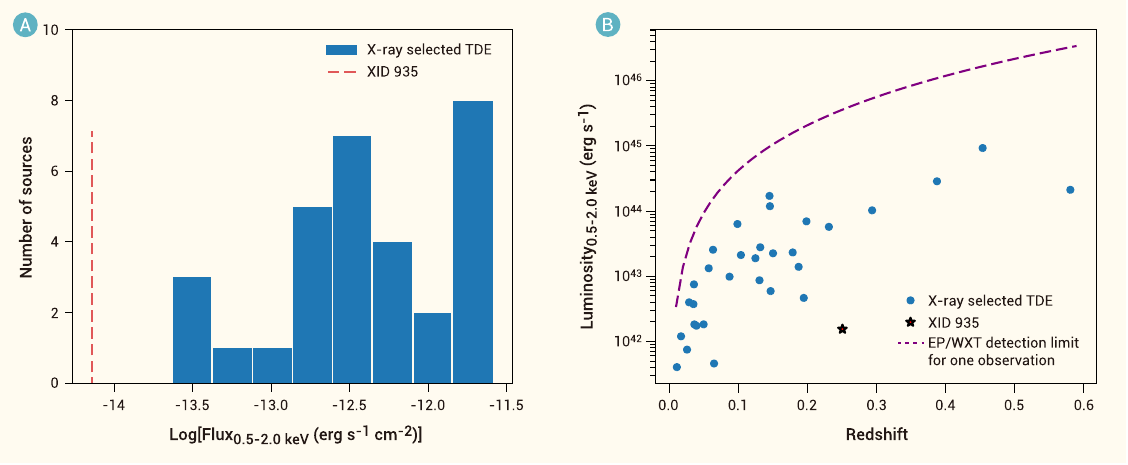}
    \caption{\textbf{X-ray-selected TDEs.} (a) Distribution of absorption-corrected peak 0.5--2.0 keV fluxes of X-ray-selected TDEs, with the vertical red dashed line highlighting XID 935. (b) Absorption-corrected peak 0.5--2.0 keV luminosities as a function of redshift, with XID 935 marked with a star symbol and other X-ray selected TDEs represented by blue points. The purple dashed curve is the 5-sigma detection limit of EP/WXT in an individual observation (i.e., 1.2-ks exposure).}
    \label{XrayTDE}
\end{figure}

Given the high level of its peak X-ray luminosity and its location at the center of its host galaxy, an edge-on disk galaxy (see Figure \ref{host.pdf}), it is reasonable to infer that the activity of XID 935 is linked to the central SMBH. As discussed above, the likelihood of XID 935 being an AGN is low. Consequently, a TDE emerges as a plausible scenario for XID 935.
Observationally, XID 935 is the observationally faintest X-ray-selected TDE candidate to date, as illustrated in Figure \ref{XrayTDE}. 
The current X-ray observational capabilities remain inadequate for detecting enough faint TDEs like XID 935 at $z > 0.1$ to enable systematic investigation of such TDEs at different redshifts. Thanks to the best X-ray coverage spanning over 20 years with the deepest total exposure of $\sim 9.5$ Ms, we can test the performance of TDE models on both long timescales and fine structural details. 

Using the $M_{\rm BH}$--$\sigma_{\star}$ relation \cite{2016ApJ...831..134V} and stellar velocity dispersion $\sigma_{\star} = 138\pm{31}\ \mathrm{km\ s}^{-1}$ measured from the Keck optical spectrum, we estimated an SMBH mass of $M_{\rm BH} = 10^{7.45\pm 0.88}\ M_{\odot}$. A non-spinning SMBH with this mass can effectively tidally disrupt a main-sequence star (see Figure 1 in Gezari 2021).\cite{gezari_tidal_2021} 

\begin{figure}[htbp]
    \centering
    \includegraphics[width=\linewidth]{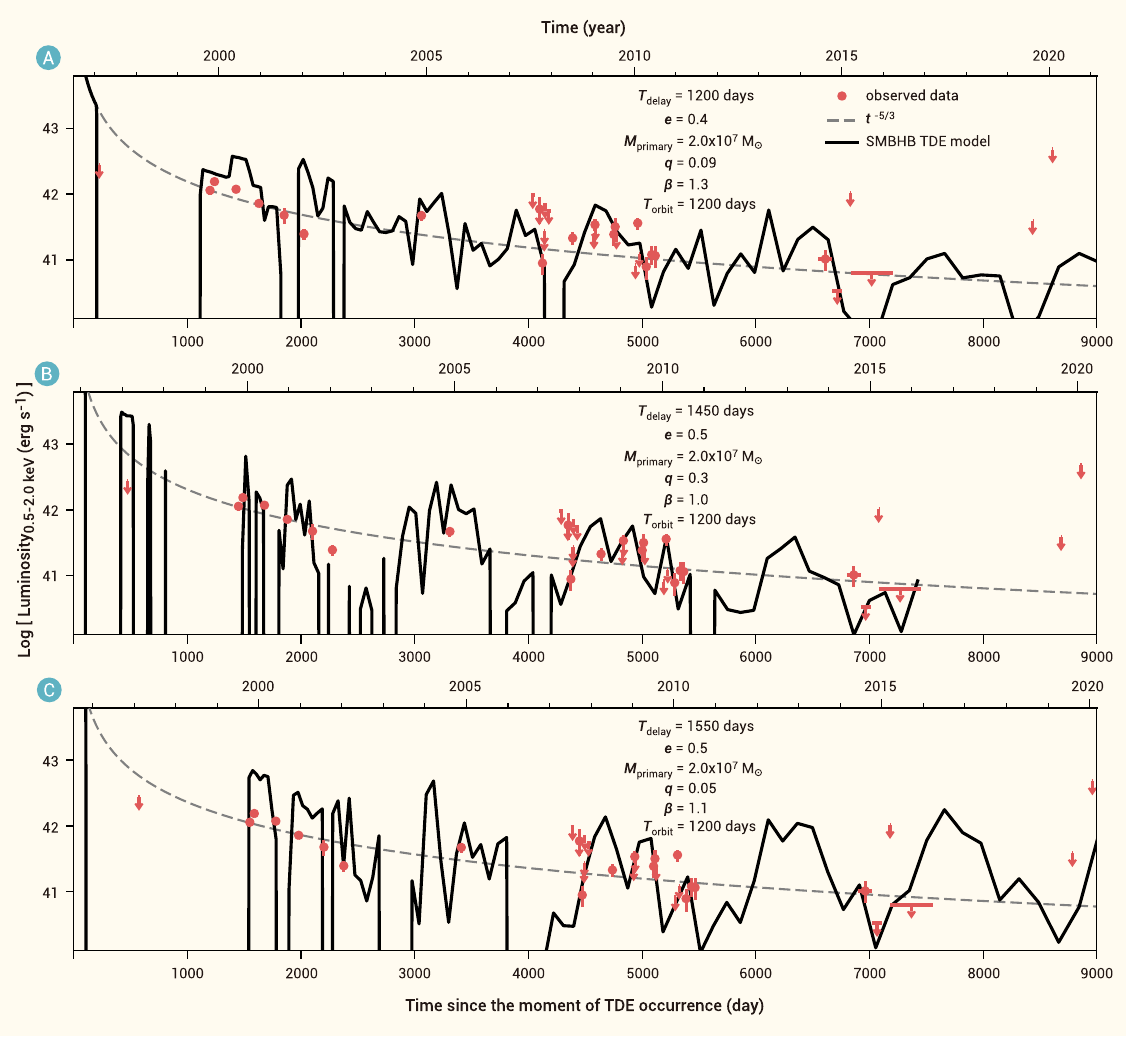}
    \caption{\textbf{SMBHB TDE model fits for XID 935.} In each panel, the rest-frame 0.5--2.0 keV light curve is shown with red symbols. Error bars correspond to 1-$\sigma$ uncertainties. Downward arrows are $90\%$-confidence-level upper limits. The dashed grey curve represents the canonical $t^{-5/3}$ decay law. The gapped black solid curve is the SMBHB TDE model fitting result, with the following model parameters annotated: $T_{\rm delay}$, the time delay between the first observed detection data point and the moment of TDE occurrence; $e$, the ellipticity of SMBHB orbit; $M_{\rm primary}$, the mass of the primary black hole; $q$, the mass ratio between the secondary and primary SMBHs; $\beta$, the penetration factor; $T_{\rm orbit}$, the orbital period of SMBHB. (A) $T_{\rm delay}=1200$ days, $e=0.4$, $q=0.09$, $\beta=1.3$. (B) $T_{\rm delay}=1450$ days, $e=0.5$, $q=0.3$, $\beta=1.0$. (C) $T_{\rm delay}=1550$ days, $e=0.5$, $q=0.05$, $\beta=1.1$. (A),(B), and (C) share identical values of $M_{\rm primary}= 2.0\times10^{7}\ M_{\odot}$ and $T_{\rm orbit}=1200$ days.}
    \label{SMBHB.pdf}
\end{figure}

The full-band X-ray luminosity of XID 935 increased by a factor of at least 27 within $\sim 2$ months, from a $90\%$ confidence level upper limit $L_{\rm 0.5-7\ keV}<10^{40.87}$ erg s$^{-1}$ (10 October 2014 -- 4 January 2015) to $L_{\rm 0.5-7\ keV}=10^{42.31\pm 0.20}$ erg s$^{-1}$ (16 March 2015). Table \ref{compare} shows the comparison between XID 935 and the two aforementioned SMBHB TDE candidates. 
\begin{table}[htbp]
\centering
\footnotesize
\setlength{\tabcolsep}{3.5pt}
\caption{Comparison between XID 935 and previous SMBHB TDE candidates}
\label{compare}
\begin{tabular}{cccc}
\hline
\textbf{} & \textbf{XID 935} & \textbf{SDSS J1201$+$30} & \textbf{OGLE16aaa}  \\ \hline
\textbf{Redshift} & 0.251 & 0.146 & 0.1655  \\
\textbf{Dimming degree} & $>11$ times (within 3 days) & $>47$ times (within 7 days) &   \\
\textbf{Flaring degree} & $>27$ times (within 70 days) &  & $>63$ times (within 9 days)  \\
\textbf{Hard X-ray emission} & Yes & No & No  \\
\textbf{Optical/UV variability} & No & No & Yes  \\ \hline
\end{tabular}
\end{table}
Remarkably, the 16 March 2015 detection showed a significance of 3.16 sigma. It is challenging for both regular TDEs and AGNs to generate such notable luminosity changes within such short timeframes,\cite{gezari_tidal_2021,2020MNRAS.498.4033T} making the 16 March 2015 detection more appropriately treated as anomalous cases in both scenarios. However, SMBHB TDEs could bring about such alterations in the accretion state due to the companion black hole perturbing the fallback rate of disrupted debris.\cite{liu_milliparsec_2014}
Indeed, the SMBHB TDE model \cite{liu_interruption_2009} can qualitatively describe the main trend of the XID 935 soft-band light curve, with three example sets of parameters being favored by the observations (see Figure \ref{SMBHB.pdf} and Section \nameref{subsec_fit} for details). 
These results show different time delays ($T_{\mathrm{delay}}$) between the first observational detection point and the moment of TDE occurrence, different ellipticities of SMBHB orbit, the same primary SMBH of $M_{\rm BH}=2.0\times 10^7\ M_\odot$, different mass ratios between the secondary and primary SMBHs, different penetration factors, and the same orbital period of the SMBHB (see Liu et al. 2014\cite{liu_milliparsec_2014} for more details about parameters.)
The relatively smooth decays observed in 2000--2002 and 2008--2010 could be reasonably explained by the regular TDE accretion stage in the SMBHB TDE model. The drop in 2015--2016 can be attributed to the perturbation of the TDE streams by the companion SMBH. However, this SMBHB TDE model is based on a simplified assumption that the bound debris of the TDE is regarded as free-moving particles, ignoring self-gravity and pressure. Hence, it cannot predict the fine structures of the light curve after several orbital periods, e.g., the fluctuations during January--July 2009 and drop during September--November 2007 (see Figure \ref{SMBHB.pdf}). A more physical SMBHB TDE model capable of explaining the fine structures of the light curve is desired. Additionally, the data gaps (i.e., no observational coverage) between 2002–2004 and 2005–2007 further complicate model validation.

The light curve of XID 935 during 1999--2002 shows a relatively smooth and steep ($\propto t^{-2.38}$) decline, suggesting a likely pTDE ($\propto t^{-9/4}$).\cite{2019ApJ...883L..17C} In contrast to complete disruption, a stellar core survives in the pTDE scenario, leading to a fallback rate that is steeper than $t^{-5/3}$. The circularization process\cite{dai_soft_2015,bonnerot_long-term_2017,2021ApJ...914...69C} seems to be a possible interpretation of the early behavior of XID 935 during 1999--2002. However, in the circularization process, even a black hole with a low mass of $M_{\rm BH} = 10^{6}\ M_{\odot}$ would not sustain a 1000-day decline. Given that XID 935 has $M_{\rm BH}\sim 10^{7.45\pm 0.88}\ M_{\odot}$, the circularization process is expected to be highly efficient, with a timescale described by
\begin{equation} \label{eq:t_cir}
t_{\rm cir} = 7\ \beta^{-4} \left(\frac{M_{\rm BH}}{10^{7.45}\ M_{\odot}}\right)^{-7/6} \left(\frac{M_*}{M_{\odot}}\right)^{-4/3} \left(\frac{R_*}{R_{\odot}}\right)^{7/2}\ {\rm days.}
\end{equation}
$t_{\rm cir}$ is given by the combination of equations (4) and (10) in Chen \& Shen (2021).\cite{2021ApJ...914...69C} Consequently, circularization is unlikely to be the primary mechanism driving the flare observed in XID 935. Nevertheless, the efficient circularization suggests that stream-stream collisions could rapidly lead to the formation of a compact accretion disk. As a result, the X-ray emission is predominantly governed by the disk accretion process. Furthermore, the hard X-ray spectrum observed during the early phase appears to originate from the accretion disk or its corona, rather than from the circularization process, which is typically associated with optical/UV emission rather than X-rays.\cite{bonnerot_first_2021}

\section*{CONCLUSIONS}

We report on a TDE candidate, XID 935, in the 7 Ms CDF-S. XID 935 has the faintest peak 0.5--2.0 keV X-ray flux among all X-ray-selected TDEs discovered to date. Thanks to the best X-ray coverage spanning over 20 years with the deepest total exposure of $\sim 9.5$ Ms, its relatively well-sampled and long-timescale light curves enabled testing of existing TDE models. 
A TDE by a single SMBH or a pTDE scenario cannot explain the primary trend of the XID 935 light curves. In contrast, the SMBHB TDE model successfully reproduces the overall behavior of the light curves; however, it still fails to match short-timescale fluctuations exactly.
Therefore, the exceptional observational features of XID 935 provide a key benchmark for refining quantitative TDE models and simulations.
Ongoing time-domain X-ray surveys conducted by facilities like the Einstein Probe \cite{2018SSPMA..48c9502Y,2025arXiv250109580J} may significantly enlarge the sample of X-ray TDEs, alongside XID 935, further enriching our knowledge of TDEs and their underlying physics.

\section*{RESOURCE AVAILABILITY}

\subsection*{Materials availability}

This study did not generate new unique materials/reagents.

\subsection*{Data and code availability}

Raw data for the observations taken with Chandra, XMM-Newton, ROSAT, and Swift are available through the HEASARC online archive services: \url{https://heasarc.gsfc.nasa.gov/docs/archive.html}. The authors can provide other data supporting the findings of this study on request.

\section*{FUNDING AND ACKNOWLEDGMENTS}

Many thanks to W. N. Brandt for his valuable comments on this manuscript.
This work is supported by the National Key R\&D Program of China (2023YFA1608100 and 2022YFF0503401), NSFC grants (12025303 and 12393814), and the Strategic Priority Research Program of the Chinese Academy of Sciences (XDB0550300). S.L. acknowledges the support by the National Science Foundation of China under grant NSFC No. 12473017. F.E.B. acknowledges support from ANID-Chile BASAL CATA FB210003, FONDECYT Regular 1241005, and Millennium Science Initiative Program – ICN12\_009. J.-H.C. acknowledges the support from the National Natural Science Foundation of China (Grant No. 12503053). 
A.V.F.'s group at UC Berkeley received financial assistance from the          
Christopher R. Redlich Fund, as well as donations from Gary and               
Cynthia Bengier, Clark and Sharon Winslow, Alan Eustace and Kathy             
Kwan, William Draper, Timothy and Melissa Draper, Briggs and Kathleen         
Wood, Sanford Robertson (W.Z. is a Bengier-Winslow-Eustace Specialist         
in Astronomy, T.G.B. is a Draper-Wood-Robertson Specialist in                 
Astronomy, Y.Y. was a Bengier-Winslow-Robertson Fellow in Astronomy),         
and numerous other donors.       
The W. M. Keck Observatory is operated as a scientific                        
partnership among the California Institute of Technology,                     
the University of California, and NASA; the observatory                        
was made possible by the generous financial support of the W. M. Keck Foundation.
The funders had no role in study design, data collection and analysis, decision to publish, or preparation of the manuscript. 

\section*{AUTHOR CONTRIBUTIONS}

Y.X. designed the project. M.H. analyzed the Chandra, XMM-Newton, and Swift data, and performed SED fitting. M.H. and Y.X. wrote the manuscript. 
S.L. and F.L. constructed the SMBHB TDE model. J.-H.C. and R.-F.S. constructed the pTDE model.
Y.Y., A.V.F., W.Z., and T.G.B. obtained the new Keck/LRIS optical spectrum. Y.W. analyzed the Keck/LRIS optical spectrum. All authors contributed to the manuscript and approved the final version.

\section*{DECLARATION OF INTERESTS}

Yongquan Xue is a member of The Innovation's scientific advisory board and was blinded from reviewing or making final decisions on the manuscript. Peer review was handled independently of this member and their research group. The other authors declare no conflicts of interest.

\section*{SUPPLEMENTAL INFORMATION}

Supplemental information includes eight figures, one listing, and six tables.

\section*{LEAD CONTACT WEBSITE}

Mengqiu Huang: https://orcid.org/0009-0003-5280-0755

Yongquan Xue: https://orcid.org/0000-0002-1935-8104

\bibliographystyle{elsarticle-num}
\bibliography{innovation}

\appendix
\includepdf[pages=-]{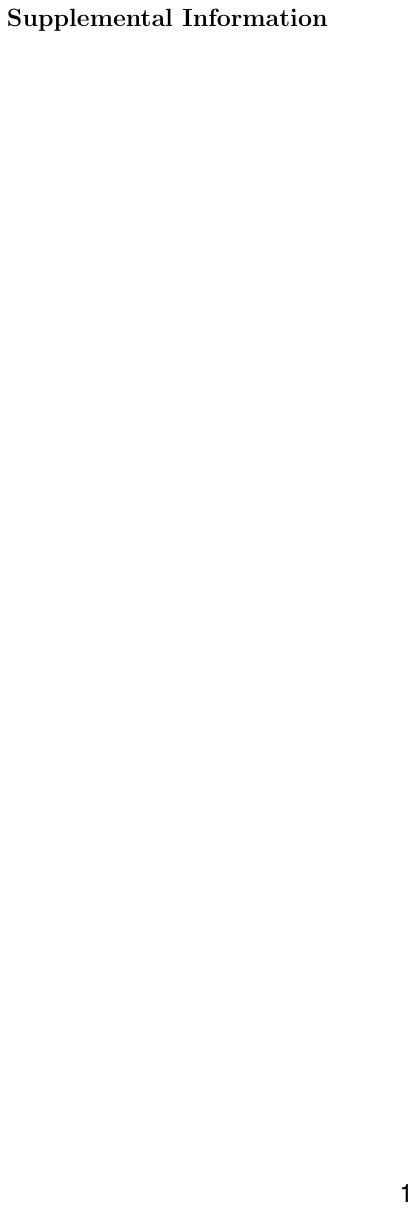}

\end{document}